# Tetrahedral collapse: a rotational toy model of simultaneous dark-matter halo, filament and wall formation

Mark C. Neyrinck[1]

[1]*Department of Physics and Astronomy, The Johns Hopkins University, Baltimore, MD 21218, USA*

22 April 2016

**ABSTRACT**
We discuss an idealized model of halo formation, in which a collapsing halo node is tetrahedral, with a filament extruding from each of its four faces, and with a wall connecting each pair of filaments. In the model, filaments generally spin when they form, and the halo spins if and only if there is some rotation in filaments. This is the simplest-possible fully three-dimensional halo collapse in the 'origami approximation,' in which voids are irrotational, and the dark-matter sheet out of which dark-matter structures form is allowed to fold in position-velocity phase space, but not stretch (i.e., it cannot vary in density along a stream). Up to an overall scaling, the four filament directions, and only three other quantities, such as filament spins, suffice to determine all of the collapse's properties: the shape, mass, and spin of the halo; the densities per unit length and spins of all filaments; and masses per unit area of the walls. If the filaments are arranged regular-tetrahedrally, filament properties obey simple laws, reminiscent of angular-momentum conservation. The model may be most useful in understanding spin correlations between neighbouring galaxies joined by filaments; these correlations would give intrinsic alignments between galaxies, essential to understand for accurate cosmological weak-lensing measurements.

## 1 INTRODUCTION

The growth of density perturbations in the Universe is well-understood in linear theory, on scales sufficiently large that the perturbations are small. Higher-order perturbation theories can predict statistics such as the power spectrum into the mildly non-linear regime, but all known schemes break down with dark-matter multi-streaming, which occurs at roughly the non-linear scale and smaller. In multi-streaming regions, the velocity field is multivalued, with dark matter on different streams having different bulk velocities at a single location.

The spherical-collapse model (Gunn & Gott 1972) allows a non-perturbative understanding of halo collapse on small scales, in a completely symmetric situation. Useful extensions have been made to the spherical collapse model, e.g. extending to ellipsoids (e.g. White & Silk 1979; Sheth et al. 2001; Angrick & Bartelmann 2010). There are also elegant self-similar spherical solutions incorporating full phase-space information (Fillmore & Goldreich 1984; Bertschinger 1985). Angular momentum (e.g. White & Zaritsky 1992; Zukin & Bertschinger 2010) and the matter accretion rate (Adhikari et al. 2014) can also usefully be added to the model. These variants of the spherical-collapse model are valuable because they are roughly correct even non-perturbatively, and, observationally, since collapsed objects are what are expected to host galaxies, the most visible extragalactic objects. These models are essential to understand important cosmological measurements, from clustering using the halo model of large-scale structure (e.g. Cooray & Sheth 2002), to cluster counting.

However, as seen in simulations, the collapse of a halo is usually sufficiently anisotropic to produce filaments and walls together with a halo, a process not included in these spherical-collapse models. Approximate large-scale-structure realizations using Lagrangian perturbation theory, such as in the Zel'dovich (1970) approximation do produce these structures, but still fail in detail after stream-crossing occurs. Lagrangian approaches generally give the morphology of the cosmic web much more accurately (compared to full simulations) than Eulerian perturbation theory (Tassev & Zaldarriaga 2012).

Structures comprising the cosmic web, like walls, filaments, and haloes, can be defined to form out of folds (Falck et al. 2012) in the dark-matter sheet, or Lagrangian submanifold (Shandarin et al. 2012; Abel et al. 2012). This approach is proving quite powerful for cosmological dark-matter simulations, as well (Hahn et al. 2013; Sousbie & Colombi 2015). These local transitions from single- to multistream regions are types of 'catastrophes.' Arnold et al. (1982) worked out a full categorization of the types of local folds that can occur in the dark-matter sheet in one and two dimensions; this was extended to three dimensions by Arnold (1982, 1983). See Hidding et al. (2014) for an exploration of these catastrophes using modern computational and visualization tools. But a halo that might be expected to host a galaxy consists of many caustics occurring together. This theory of catastrophes does not capture the complexity of a full collapsed structure. Here we develop a theory which does, albeit in a toy model.

As explored in many of the above works, the cosmic web forms in analogy with the origami-folding of a paper sheet (e.g. Falck et al. 2012; Neyrinck 2012; Neyrinck et al. 2015). In paper origami, a 2D sheet that cannot stretch is folded in the usual three dimensions. In cosmology, the sheet is a stretchable 3D manifold, which folds up in 6D position-velocity phase space. As in paper origami, the sheet is continuous and cannot tear, and cannot pass through itself in 6D. This concept is already essentially present in a Lagrangian fluid-dynamics viewpoint (following mass elements, not fixed locations). If the dark matter were not collisionless, Lagrangian (initial-conditions) patches could not pass through each other in position space; instead, physical shocks of gas would form, and cosmic web components would be much different.

Recently, I (Neyrinck 2015a,b) investigated the consequences of taking this origami interpretation rather literally, in an 'origami approximation.' It imposes a strong assumption: that the dark-matter sheet does not stretch, and the Lagrangian-to-Eulerian (initial-to-final conditions) mapping $\bm{q} \to \bm{q} + \bm{\Psi}(\bm{q})$ is a continuous piecewise isometry. That is, the density on each stream is constant (and undefined at caustics). Density variations in Eulerian space can therefore only arise through multistreaming, i.e. through caustic formation.

The no-stretching assumption is, in general, wrong. Indeed, in the Zel'dovich approximation, structures are built entirely from such 'stretching.' However, the no-stretching assumption is not as bad as one might at first think. In full gravity, it seems that the density on each stream seems to vary remarkably little, something that remains to be explained. Vogelsberger & White (2011) found that even in high-resolution, cold-dark-matter Aquarius (Springel et al. 2008) haloes, median fine-grained densities on streams vary only by about an order of magnitude with distance from halo centers, whereas the total density (summed over streams) increases by a factor of a million in the center. Also, Hahn et al. (2014) recently measured the velocity divergence (giving the stretching rate of sheet layers overlapping at a position) in a warm-dark-matter simulation, using a phase-space-sheet method giving unprecedented accuracy. They found that the velocity divergence is remarkably uniform spatially, predominantly positive even in multistream regions (except at caustics).

A second, entirely reasonable assumption in the origami approximation is that void regions are irrotational. This should hold in reality, since expansion quickly dampens any primordial vorticity. As Pichon & Bernardeau (1999) showed, stream-crossing even in the potential-flow Zel'dovich approximation produces vorticity in multi-stream regions, but not in single-stream regions. Vorticity is indeed closely tied to stream crossing, in full simulations (Hahn et al. 2014; Wang et al. 2014). The origami approximation gives an idealized cosmic web, with convex polyhedral nodes and voids, planar walls, and filaments that are polygonal in cross-section. Also, in this approximation, voids do not move much, compared to other elements; this has some physical validity, since large void centers correspond to physical potential maxima, thought to move little compared to other elements in the cosmic web.

A halo's or galaxy's spin seems to depend on its cosmic-web environment. This has been found in simulations (Aubert et al. 2004; Aragón-Calvo et al. 2007; Hahn et al. 2007; Sousbie et al. 2008; Paz et al. 2008; Zhang et al. 2009; Codis et al. 2012; Libeskind et al. 2013; Aragon-Calvo & Yang 2014) and even in observations (Tempel et al. 2015; Zhang et al. 2015). Even if one is uninterested in cosmic-web dynamics, this is an important effect to understand for cosmological measurements, because it would produce 'intrinsic alignments' of galaxy ellipticities, and therefore a systematic error in gravitational weak-lensing measurements (e.g. Codis et al. 2015a). Weak lensing promises to be a powerful method to survey the matter density field, through the correlations between galaxy ellipticity directions that gravitational lensing would give. But any intrinsic galaxy shape correlations would contaminate this measurement.

Halo spin is usually understood within the context of the tidal torque theory (for a review, see Schäfer 2009), in which Lagrangian patches are spun up because of a misalignment, and therefore torque, that exists between the inertia and tidal tensor in a general situation. Recently, Codis et al. (2015b) developed a theoretical formalism that explains the different types of halo-filament spin alignments within the tidal torque theory.

Observationally, spin correlations between spiral galaxies neighbouring each other (with distance $\lesssim 1h^{-1}$ Mpc) have been measured (Slosar et al. 2009; Lee 2011), although the statistical significance of the detections may be overestimated (Andrae & Jahnke 2011). The tidal torque theory is consistent with the correlation, but it does not fully explain the correlation at short distances.

Given the observational relevance of galaxy spins, we pay special attention in this paper to halo spins and how they relate to filaments that form concurrently with the halo. In the origami approximation, the important property of halo spin naturally comes out of a node collapse model, and is related to the rotations of filaments. Because of the rigidity of the approximation, we do not expect its predictions to hold precisely, but because haloes in simulations seem to form concurrently with cosmic-web elements around them, just as happens in the toy model, we expect its predictions to provide a guide for intuition qualitatively, and perhaps quantitatively as well, if properly calibrated.The model could be particularly applicable to the first stages of halo formation, before virialization. Afterward, it is likely that many quantities will remain relevant, such as filament orientations and spins, and the node's spin.

In the paper, first, we discuss collapses in 1D and 2D, then how velocities relate to them. In the bulk of the paper, we then discuss three-dimensional (tetrahedral) collapses, with regular and then irregular filament arrangements, giving examples and simple relationships.

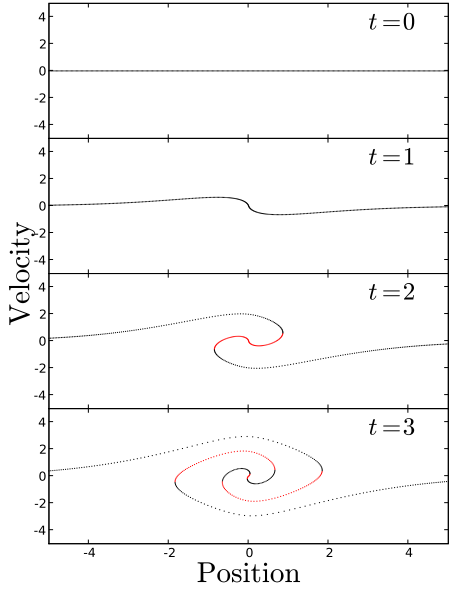

**Figure 1.** A schematic phase-space spiral that occurs in a 1D collapse of the dark-matter 'string' (sheet in > 1D). Vertices (particles) on this 1D mesh are represented as dots. Caustics, or creases, occur where the string goes vertical. Black particles are oriented as in the initial conditions; red particles are oriented with opposite parity. Note that here, the dark-matter string stretches, i.e. the particles vary in their distances. But, the same pattern of creases can be produced without stretching. The times $t = 0, 1, 2, 3$ represent stages in the collapse, likely not equally-spaced physical times.

## 2 1D AND 2D ORIGAMI-APPROXIMATION COLLAPSE

As usual, it is useful to build understanding by considering simpler, lower-dimensional cases. This is especially so here, where many elements are built from extrusions of lower-dimensional elements. Also, effectively 1D and 2D collapses are relevant even in 3D, where haloes can form within pre-existing walls and filaments, effectively 2D and 1D universes.

In 1D, collapsed structures are delineated by their outer caustics; for example, as in Fig. 1, the canonical phase-space spiral in 2D phase space. From here on, we focus on the outer caustics of structures, by looking at 'simple' collapsed regions, i.e. without inner phase-space windings. A 1D simple node consists of just two caustics, that separate a three-stream region from the single-stream background.

In 2D, the simplest collapsed structure is a *filament*, an extrusion of a 1D node. All caustics must be straight lines due to the piecewise isometry condition (Demaine & O'Rourke 2008). Furthermore, the caustics must be parallel lines (the origami term for this is a *pleat*). This is because, if caustics were non-parallel, the two reflections they produce would cause neighbouring voids to be rotated with respect to each other. Again, we consider 'simple' filaments consisting of just two caustics, but inner caustics could be added inside a simple filament.

'Polygonal collapse' occurs at the intersection of filaments, at a convex-polygonal *node*. It is convex because Kawasaki's theorem (Kawasaki 1989b) (both alternating sums of vertex angles add to 180°), applied to a vertex of a polygon, prescribes that all angles between caustics are < 180°. Each vertex joins an even number ⩾ 4 of caustics. So, nodes cannot form in isolation; they must form together with e.g. filaments. This can be thought of in terms of gathering together a patch of paper, which cannot be done without making creases that radiate away from the patch (i.e., filaments). Kawasaki's theorem, with some simple geometry, also implies that angles at which all filaments come off of a node's edges must be equal (e.g. Kawasaki 1997). The Lagrangian-to-Eulerian mapping gives a rotation of the node, producing filaments at the same time.

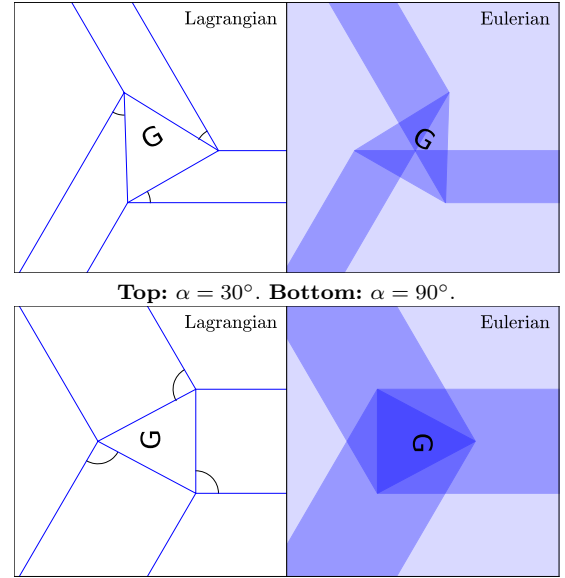

**Figure 2.** Triangular-collapse models, with different rotation angles $\alpha$. The bottom panels show irrotational triangular collapse, the closest triangular analog to circular collapse that an unstretchable dark-matter sheet allows.

Circular (2D spherical) collapse is impossible without stretching the dark-matter sheet, because caustics must be straight lines. Circular collapse is one special case of the collapse of a region, with isotropy around the node, and in which the sheet stretches substantially. Polygonal collapse can be seen as another special case, with anisotropy, but no stretching. However, it must be noted that most polygonal collapse situations likely do not correspond to an actual dynamical gravitational collapse.

Fig. 2 shows two examples of triangular collapse. At left are the caustics which produce these collapses in Lagrangian space. The origami term for this is a *crease pattern*; each line is a caustic that effects a reflection in the Lagrangian-to-Eulerian mapping. In both, the collapsing polygon is an equilateral triangle, but they differ in the angle $\alpha$ (indicated by arcs) with which the filament caustics come off of the triangle's sides. From Lagrangian to Eulerian space, the triangle rotates by $2\alpha$. The bottom panels show irrotational triangular collapse, which can be considered a parity inversion of all elements, since a rotation by 180° is a 2D reflection. Irrotational ($\alpha = 90°$) collapse is the closest analog to spherical collapse in the origami approximation.

The topological and global properties of a cosmic web in the origami approximation are worth mentioning. Both haloes (in Lagrangian space) and voids are convex polygons (polyhedra in 3D), completely enclosed by multistream walls and filaments. Walls, too, are convex-polygonal in 3D. While this does correspond to an idealized, but qualitatively accurate cosmic web in the density field, somewhat like the Voronoi model of (Icke & van de Weygaert 1991), we note that voids do not generally seem to be such idealized structures in simulations, entirely enclosed by stream crossings (Falck & Neyrinck 2015).

## 2.1 Complexity of 2D nodes

How reasonable is it for us to restrict attention to simple nodes, in which exactly 1 polygon collapses? What 'complex' nodes (with $> 1$ polygon) exist? A more precise definition would be useful to answer this question. *Node-polygons* are bounded polygons that all border each other, such that a path along a curve encircling all of them encounters only filaments. That is, the path encounters regions alternating between filament regions bordered by parallel lines (these undergo a reflection), and void regions which undergo no reflections, rotations, or parity inversions. The node itself is the union of these polygons.

A full categorization of simple and complex nodes would be interesting to undertake, but for now, we just describe a few examples of complex nodes. First we describe nodes from which only 3 filaments emerge; this is the most cosmologically relevant case. There is no node with 2 polygons. This is because each polygon would have positive parity, since it must border a negative-parity filament region. But the 2 polygons must also border each other, meaning they have different parity; this cannot happen.

There do exist nodes with 3 node-polygons and 3 filaments coming off; an example is shown at left in Fig. 3. As indicated by the two angles $\alpha_1$ and $\alpha_2$, this node is essentially a combination of two simple nodes, each with its own shape and angle of twisting. Another solution, with 7 node-polygons, is shown in the right panel of Fig. 3. This node is another kind of a combination of two simple twist folds. Inner twist folds can be added *ad infinitem*; it would be interesting to see if such models have any relation to the inner structure of realistic haloes. A different way to combine nodes is to explicitly concatenate them, taking the outcome of one collapse as the initial conditions for the next. These would give filaments with inner structure, as well.

We show one more model with 3 node-polygons in Fig. 4, with 4 filaments coming off of it. This, too, is a sort of combination of 2 simple nodes (the two outer triangles).

In conclusion, while there exist 'complex' 2D collapse models, many of them are combinations of 'simple' one-polygon models, and thus it makes sense to concentrate attention on the simple model. We suspect the same kind of statement applies to 3D, as well, something for future investigation.

## 3 KINEMATIC MODELS

These origami models give a direct initial-to-final (Lagrangian to Eulerian) mapping. However, reducing the models' applicability to reality, they are not kinematic. Explicit velocities are also important to check the paper-origami-like requirement that the dark matter sheet never cross itself in phase space. This requirement exists in the cosmological case because in a dissipationless and collisionless dynamical system, particles cannot have different phase-space coordinates initially, but have the same coordinates later. The no-crossing condition would be violated if at some position, velocities were the same on any two streams. For all cases investigated here, the velocities differ on each stream; they are different interpolations of different linear transformations and translations, coming from different Lagrangian positions. However, it remains to be shown that the no-crossing condition always holds in such collapses.

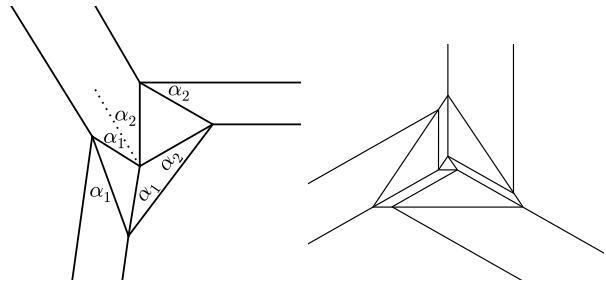

**Figure 3.** Two complex nodes that each produce 3 filaments. Only the locations of the caustics in Lagrangian space are shown; in origami language, only the crease patterns are shown, but no folded forms. They are different types of combinations of simple nodes. **Left**: An example of the 'simplest' complex node, with only 3 polygons. The dotted line is not a caustic or crease, but is drawn to clarify the relation of this node to a pair of simple nodes with twist angles $\alpha_1$ and $\alpha_2$. **Right**: A node with 7 polygons, this time a combination of one simple node inside another.

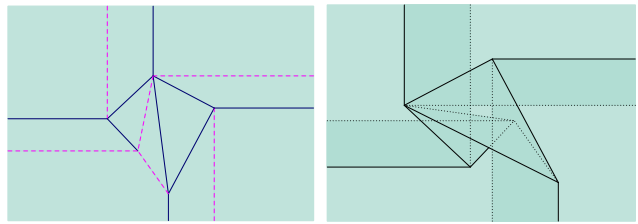

**Figure 4.** A complex node (twist fold, in origami) with 3 node-polygons and 4 filaments. **Left**: Origami crease pattern (Lagrangian space), with mountain folds (pointing upward if viewed from the side) shown in solid black, and valley folds (pointing downward if viewed from the side) shown in dashed lavender. **Right**: Origami folded form (Eulerian space). Figure courtesy Robert Lang.

Two ways come to mind to make the models kinematic, i.e. to define velocities. The first way is to associate a time variable with a parameter indicating the degree of folding progress, and track the velocity of each patch of the sheet as it folds. Without sheet-stretching, for each stream (Lagrangian region bordered by caustics), there is an orthogonal transformation composed with a translation that constitutes the Lagrangian-to-Eulerian mapping $\boldsymbol{q} \rightarrow \boldsymbol{\Psi}(\boldsymbol{q}) + \boldsymbol{q}$. This mapping can be done incrementally, interpolating between the identity mapping and the full displacement. There is an inconsistency in this approach: applying a transformation partially would generally not give a model satisfying the no-stretching condition. But there is no physical reason to enforce the no-stretching condition, anyway. Interpolating the $\mathbf{\Psi}$ mapping seems to be a good strategy for future investigations.

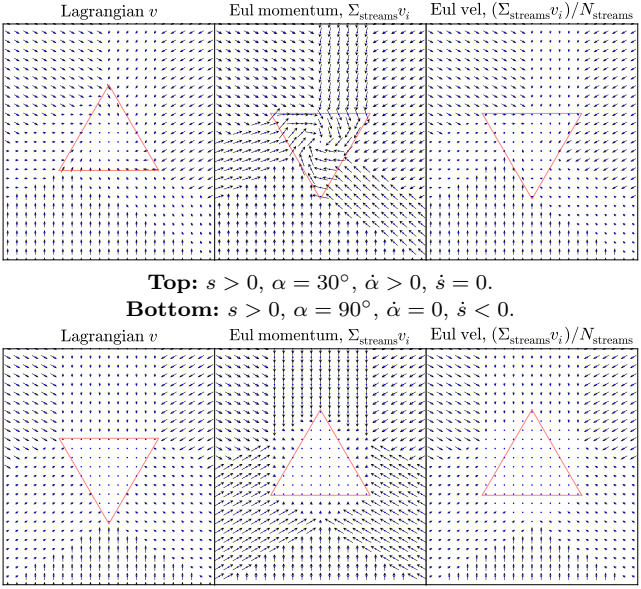

**Figure 5.** Kinematic triangular-collapse models, with different values of $\alpha$ (the angle) and $s$ (a scale parameter), and their time-derivatives. **Left**: Initial state. **Middle, Right**: final state, showing both momentum and mean velocity. **Top**: purely rotational collapse, in which only $s$ increases with time. **Bottom**: irrotational collapse, where only $s$ increases with time. At the snapshots shown, these correspond (if tilted by 90°) to the models in Fig. 2.

There is a second way to introduce velocities: enforce no-stretching at all times, but continuously deform the model, i.e. the locations of caustics in Lagrangian space. Some exploration of this approach for a 2D model appears in Neyrinck (2015b), which we essentially repeat here. A polygonal-collapse model with fixed shape has two parameters: $s$, the scale of the model (e.g. a side length); and $\alpha$, the angle. Velocities may be defined from $s(t)$ and $\alpha(t)$. In this approach, densities are simply proportional to the number of streams, since each stream is constant-density.

Fig. 5 shows two polygonal-collapse models with different sets of parameters: purely rotational, and purely irrotational models. As expected, there is obvious vorticity in the rotational model in the node, but not elsewhere. Curiously, the mean velocities (right panels) in multistream regions are lower than in the voids. In the filaments, this is because the velocity is the average over three streams, two of which have rather small components pointing toward the node. But as expected, the Eulerian momentum is large.

These velocity fields might be of use in identifying features in the density and velocity fields corresponding to the positions of a halo's outer caustic (Falck et al. 2012), or 'splashback radius' (Diemer & Kravtsov 2014; More et al. 2015, 2016). Delineating the edges of a collapsed region can be easily done using the ORIGAMI algorithm (Falck et al. 2012) if the displacement field is known, but identifying features in the velocity and density fields could make this definition more directly useful in observations.

# 4 3D ORIGAMI-APPROXIMATION COLLAPSE

Now we turn attention to 3D. Just as in 2D, a 3D node cannot form in isolation, in the origami approximation. If the dark-matter sheet folds flat (i.e., fits into a 3-dimensional space) with no stretching, there is a condition on vertices that join together caustics (Robertson 1978; Kawasaki 1989a; Hull 2010), which we call 'Kawasaki's theorem in 3D.' In 3D, caustics are 2D surfaces. The regions tiling an arbitrarily small sphere centered at a vertex divide into equal numbers of regions that end up with positive (the same as in the initial conditions) and negative parity. The condition for flat-folding is that the solid angles of each parity sum to $2\pi$ steradians. Since each solid angle meeting at a vertex $< 2\pi$, and the caustic surfaces are flat (by no-stretching), the central node must be a convex polyhedron, with additional caustics coming off. As in 2D, the lack of mutual rotation in voids separated by a wall requires the wall caustics to be parallel; the simplest way these can join together is, in cross section, the same as a 2D node. That is, a filament sprouting walls in 3D is an extrusion of a 2D polygonal node sprouting filaments.

3D collapses that are essentially lower-dimensional are possible, such as 2D triangular collapse in an already-collapsed wall. This would lead to the apparently common three-filament haloes (Dekel et al. 2009; Danovich et al. 2012). The simplest truly 3D collapse is of a polyhedron with the lowest number of sides, i.e. a tetrahedron. Polyhedral collapses with more sides are possible, but haloes in simulations seldom have more than 4 obvious filaments. Indeed, an unlikely, special arrangement of void centers (in a Voronoi model) would be necessary to produce a node joining more than 4 voids. We expect that going to a higher number of filaments would not increase the agreement of the polyhedral collapse model with simulations.

## 4.1 Method of solution

The properties of the tetrahedral-collapse model that may have some relation to observations are the filaments' directions, spins, and masses per unit length, as well as the mass, spin, and (perhaps) shape of the central node. Detailed shapes of the various components will almost certainly virialize away after protohalo collapse, although some aspects of a node's shape, e.g. its ellipticity, may persist.

We wish to determine what minimal set of observationally-relevant properties determine all properties of the tetrahedral-collapse system. First we will enumerate the DoF of various components of the model. A tetrahedron has 12 total DoF: 3 coordinates for each of 4 vertices. These can be separated into 3 translational DoF, 1 scale, 5 DoF specifying the shape (Rassat & Fowler 2004), and 3 DoF of a 3D rotation.

5 parameters specify the 4 filament orientations in 3D, after choosing a set of axes. Let $\hat{\boldsymbol{f}}_i$ denote the unit vector giving the direction of the $i$th filament. Choose the $z$-axis to align with filament 1, and the $x$-axis to point along filament 2 as seen looking down filament 1. So $\hat{\boldsymbol{f}}_1 = (0, 0, 1)$, and $\hat{\boldsymbol{f}}_i = (\sin\theta_i \cos\phi_i, \sin\theta_i \sin\phi_i, \cos\theta_i)$ for $i$ from 2 to 4. $\phi_2 = 0$, from the alignment of filament 2 with the $x$-axis, giving a total of 5 nonzero angles $\theta_i$ and $\phi_i$.

6     *Mark C. Neyrinck*

As we find below, the 5 parameters of the filament arrangement, along with 3 additional parameters (specified here as 3 of the 4 filament rotations), fully specify all relevant quantities of the collapse. The 'relevant' quantities here are the 4 filament rotations, shapes, and relative cross-sectional areas; the 6 relative wall thicknesses; and the shape and rotation of the node. Another choice of the 3 DoF (e.g., the 3D rotation of the node from initial to final conditions) would likely specify the system, as well. ('Irrelevant' quantities are translations and rotations of the whole system, and an overall scaling parameter.)

The three additional inputs into the model (beyond the filament arrangement) are filament rotation half-angles $\alpha_i$, which run from 0 to $\pi$. (The full rotation from Lagrangian to Eulerian space is $2\alpha$). If $\alpha < \pi/2$, the rotation is in one direction, and if $\alpha > \pi/2$, it is in the other. If $\alpha = \pi/2$, there is no rotation; the motion of the filament is a pure parity inversion, about its axis. A rotation by an angle $\alpha_1$ can also be thought of as a rotation by $\pi - \alpha_1$ in the opposite direction.

We solve for the 4 node vertex positions, which we call $g_i$, in a somewhat brute-force approach, relying on Mathematica to solve systems of equations numerically.

Let $\boldsymbol{x}_{\perp i} \equiv \boldsymbol{x} - \hat{\boldsymbol{f}}_i(\hat{\boldsymbol{f}}_i \cdot \boldsymbol{x})$ denote the component of the vector $\boldsymbol{x}$ perpendicular to $\hat{\boldsymbol{f}}_i$. Set $\boldsymbol{g}_4 = 0$; this removes the translational DoF. The $x$ and $y$ coordinates of $\boldsymbol{g}_2$ and $\boldsymbol{g}_3$ are easy to get, from our choice of $\hat{\boldsymbol{f}}_1$ along the $z$-axis. With reference to Fig. 6, $\boldsymbol{g}_{3\perp 1}$ and $\boldsymbol{g}_{2\perp 1}$ (' $\perp 1$' fixes attention on the $x$ and $y$ coordinates) can be put in terms of $\phi_3$, $\phi_4$, and a scale parameter, $s_{34\perp 1} \equiv |(\boldsymbol{g}_3 - \boldsymbol{g}_4)_{\perp 1}|$, which we set to unity. 5 parameters remain (the $z$-coordinates of $\boldsymbol{g}_2$ and $\boldsymbol{g}_3$, and all 3 coordinates of $\boldsymbol{g}_1$). 6 constraints come from requiring each of 3 walls to come off of filaments 2 and 3 at the correct angles, $\alpha_2$ and $\alpha_3$. These constraints are of the form $\hat{\boldsymbol{f}}_{a\perp b}\hat{\cdot}(\boldsymbol{g}_c - \boldsymbol{g}_d)_{\perp b} = \cos\alpha_b$, where $a,b,c,d$ run from 1 to 4, and are all different. The shorthand $\hat{\cdot}$ means a normalized dot product, returning the cosine of the angle between its arguments, i.e., $\boldsymbol{a}\hat{\cdot}\boldsymbol{b} \equiv (\boldsymbol{a}\cdot\boldsymbol{b})/(ab)$. The 6 constraints are:

$$\hat{\boldsymbol{f}}_{1\perp 2}\hat{\cdot}(\boldsymbol{g}_4 - \boldsymbol{g}_3)_{\perp 2} = \cos\alpha_2, \quad \hat{\boldsymbol{f}}_{3\perp 2}\hat{\cdot}(\boldsymbol{g}_1 - \boldsymbol{g}_4)_{\perp 2} = \cos\alpha_2,$$
$$\hat{\boldsymbol{f}}_{4\perp 2}\hat{\cdot}(\boldsymbol{g}_3 - \boldsymbol{g}_1)_{\perp 2} = \cos\alpha_2, \quad \hat{\boldsymbol{f}}_{2\perp 3}\hat{\cdot}(\boldsymbol{g}_4 - \boldsymbol{g}_1)_{\perp 3} = \cos\alpha_3,$$
$$\hat{\boldsymbol{f}}_{4\perp 3}\hat{\cdot}(\boldsymbol{g}_1 - \boldsymbol{g}_2)_{\perp 3} = \cos\alpha_3, \quad \hat{\boldsymbol{f}}_{1\perp 3}\hat{\cdot}(\boldsymbol{g}_2 - \boldsymbol{g}_4)_{\perp 3} = \cos\alpha_3. \quad (1)$$

The constraints around filament 4 are not used for solving for $\boldsymbol{g}_i$, but we do check that the $\boldsymbol{g}_i$ satisfy them:

$$\hat{\boldsymbol{f}}_{1\perp 4}\hat{\cdot}(\boldsymbol{g}_3 - \boldsymbol{g}_2)_{\perp 4} = \cos\alpha_4, \quad \hat{\boldsymbol{f}}_{2\perp 4}\hat{\cdot}(\boldsymbol{g}_1 - \boldsymbol{g}_3)_{\perp 4} = \cos\alpha_4,$$
$$\hat{\boldsymbol{f}}_{3\perp 4}\hat{\cdot}(\boldsymbol{g}_2 - \boldsymbol{g}_1)_{\perp 4} = \cos\alpha_4. \quad (2)$$

We solve for the 5 unknowns numerically, using Mathematica's FINDROOT function. First, we try we solving for them from the first 5 of Eqs. (1). The initial guesses are the unknowns' values for $\alpha_i = \pi/2$, and regular-tetrahedral filament directions. Occasionally, solving the first 5 equations gives solutions that do not satisfy the sixth; then, we try excluding the fourth or fifth instead of the sixth from the set of equations fed to FINDROOT. At least one of these seems to always give a unique solution that satisfies all 6 equations. The remaining angle $\alpha_4$ can then be found from one of Eqs. (2). Somewhat remarkably, all cases we tried gave an $\alpha_4$ satisfying all Eqs. (2), i.e. a fourth filament that is also a twist-fold, or 2D collapse, in cross-section.

For the folded-up model, the only difference is that the $\alpha_i$'s switch sign; going from Lagrangian to Eulerian space, $\alpha_i \to -\alpha_i$. The initial and final positions determine the rotation matrix **M**, such that $\boldsymbol{g}_i \to -\mathbf{M}\boldsymbol{g}_i$ gives the transformation from initial to final. In all cases tried, the linear transformation that the $\boldsymbol{g}_i$ undergo is an 'improper rotation'; the node is inverted through the origin, and then rotated. The rotation angle $\omega$ can be found using the formula $2\cos(2\omega) = \text{Tr}(\mathbf{M}) - 1$. The factor of 2 is inside the cosine to conform with our convention that $\alpha_i$ is the angle between caustics in Lagrangian space; it is half of the full angle by which filaments rotate from Lagrangian to Eulerian space. The direction of the vector $\boldsymbol{\omega}$ is defined to be the axis along which the node rotates. This vector may be found since it is is the only real eigenvector of **M**, with eigenvalue 1.

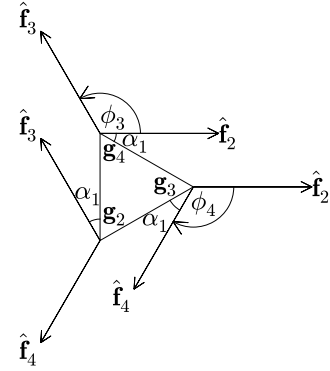

**Figure 6.** A cross-section through filament 1, looking down its barrel from the $+z$-axis. All vectors should have a '$_\perp 1$' subscript, denoting the component of the vector perpendicular to $\hat{\boldsymbol{f}}_1$. The $x$ and $y$ axes are oriented as usual in the page, and the $z$ axis points out of the page, aligned with $\hat{\boldsymbol{f}}_1$. The angle labeled $\phi_3$ is so labeled for clarity; in fact $\phi_3$ is its complement, and the labeled angle is $2\pi - \phi_3$.

In cross-section, by their construction as 2D twist folds, the angles of filaments in the model obey Kawasaki's theorem in 2D (giving conditions for the 2D model to 'flat fold' back into a plane). But it is also important to verify that Kawasaki's theorem in 3D is obeyed at each vertex (checking the condition for the 3D model to 'flat fold' back into a 3D manifold). Solid angles around a vertex can be separated into two sets of solid angles, with opposite parity. The condition is that the sums of solid angles of both parities equal $2\pi$ steradians. In polyhedral collapse, voids, walls, filaments, and the node have parities 1, -1, 1, and -1, respectively. Around vertex $\boldsymbol{g}_4$, the condition is that, summing solid angles subtended by all regions of positive parity (first, the void region, and then the three filaments),

$$\Omega(\hat{\boldsymbol{f}}_1, \hat{\boldsymbol{f}}_2, \hat{\boldsymbol{f}}_3) + \Omega(\boldsymbol{g}_2 - \boldsymbol{g}_4, \boldsymbol{g}_3 - \boldsymbol{g}_4, \hat{\boldsymbol{f}}_1) +$$
$$\Omega(\boldsymbol{g}_3 - \boldsymbol{g}_4, \boldsymbol{g}_1 - \boldsymbol{g}_4, \hat{\boldsymbol{f}}_2) + \Omega(\boldsymbol{g}_1 - \boldsymbol{g}_4, \boldsymbol{g}_2 - \boldsymbol{g}_4, \hat{\boldsymbol{f}}_3) = 2\pi, \quad (3)$$

where $\Omega(\boldsymbol{a}, \boldsymbol{b}, \boldsymbol{c})$ gives the solid angle subtended by a triangle with vertices along vectors $\boldsymbol{a}$, $\boldsymbol{b}$, and $\boldsymbol{c}$ (Van Oosterom & Strackee 1983):

$$\Omega(\boldsymbol{a}, \boldsymbol{b}, \boldsymbol{c}) = 2\tan^{-1}\frac{|\boldsymbol{a}\cdot(\boldsymbol{b}\times\boldsymbol{c})|}{abc + (\boldsymbol{a}\cdot\boldsymbol{b})c + (\boldsymbol{a}\cdot\boldsymbol{c})b + (\boldsymbol{b}\cdot\boldsymbol{c})a}. \quad (4)$$

Equivalently, we could have summed the solid angles in the regions of negative parity: the node and the walls. In all cases tried, the 3D version of Kawasaki's theorem was satisfied at all vertices.

Readers interested in full details of the solution method may wish to view the interactive Wolfram Computable Document Format (CDF) file, containing Mathematica code for the solution, at http://skysrv.pha.jhu.edu/~neyrinck/TetCollapseFil2Node/.

### 4.2 Results: Regular-tetrahedral collapse

Having described the method of solution, we now describe some results and properties of solutions. First, we investigate 'regular-tetrahedral collapse,' with filament angles $\hat{\boldsymbol{f}}_i$ pointing toward vertices of a regular tetrahedron (this does not generally imply that the node is also a regular tetrahedron). A few simple, interesting relationships hold among the various angles in the regular case, which we found empirically. This implies that there is a more elegant method of solution.

First, we found an angular momentum-conservation-like law: if $A_i$ is the cross-sectional area of filament $i$, and $\alpha_i$ is the angle at which walls come off of the filament in Lagrangian space,

$$\sum_{i=1}^{4} A_i \sin(2\alpha_i) = 0. \quad (5)$$

A filament with $\alpha_i = 0$ or $\pi$ does not collapse (twist) at all, so it makes sense that such a filament does not contribute to a sum involving rotation. Also, zero contribution with $\alpha = \pi/2$ makes sense, since a rotation by $2\alpha = \pi$ is the same as a pure parity inversion, with no rotation.

Physically, the filament's quantity $A_i \sin(2\alpha_i)$ looks similar to an angular momentum (per unit length): in Lagrangian space, $A_i$ is a mass per unit length, and $2\alpha$ is the angle by which the filament rotates. However, a true angular momentum would scale with $A_i^2$ if the filament stays fixed in size, since both the mass of a filament and its mean-square displacement from the axis scale with $A_i$. One interpretation of $\sum_i A_i \sin(2\alpha_i) = 0$ would be as the sum of angular momenta of filaments after they have all contracted to the same radius, in Eulerian space. It is intriguing that in the filaments there is a kind of conservation of angular momentum, which was zero in the initial conditions.

Another relationship involves a constant for all filaments around a node (this constant is generally different for different tetrahedral collapses):

$$A_i \sin^2 \alpha_i = \text{const}, \forall i \quad (6)$$

Eqs. (5) and (6) can be combined with the help of trigonometric identities to give the following equation relating only the angles of each filament's rotation:

$$\sum_{i=1}^{4} \cot \alpha_i = 0. \quad (7)$$

So, three angles $\alpha_i$ and an area $A_i$ suffice to determine the other $\alpha_i$ and the three remaining $A_i$'s. Two values of the unknown angle satisfy Eq. (7), because of the cotangent function's period $\pi/2$, but only one of those will satisfy Eq. (5).

The following illustrations show the locations of caustic surfaces before and after collapse, in two simple examples. Using a 3D-capable viewer (Adobe Reader) is recommended.

But a 2D view down filament 4 is still shown in a non-3D-capable viewer (or before clicking on figures in a 3D-capable viewer). The caustics in the 2D view are colour-coded to indicate caustics forming walls (blue), filaments (green), and the node (yellow and red).

While interacting with these 3D figures will give a rather good sense of the geometry of the outer caustic surfaces, they likely leave mysterious the motions of the caustics from their Lagrangian to Eulerian locations. Animations showing the motion constituting the collapse (and also giving some sense of the geometry, particularly useful for those without a 3D-capable viewer) are provided in the Supplementary Material, also linked from each figure caption. But the best way to explore the model is by interacting with the Wolfram CDF file linked above.

Figs. 7 and 8 shows the simplest, most symmetric model, with $(\alpha_1, \alpha_2, \alpha_3, \alpha_4) = (\pi/2, \pi/2, \pi/2, \pi/2)$. This is an irrotational model, in which filaments rotate by an angle $\pi$, which is the same as a parity inversion through the filament's axis. In the 2D views, the parity inversion can be observed in the node, as well; a single face of the tetrahedron is visible in the Lagrangian view, but three (others) are visible in the Eulerian view. In this irrotational model, after folding, voids have 1 stream, walls have 3 streams, the centers of filaments have 7 streams, and the node is where all 15 Lagrangian-space streams overlap. These numbers may be useful for cosmic-web characterizations based on stream number (Shandarin & Medvedev 2014; Ramachandra & Shandarin 2015).

Figs. 7 and 9 show the only other solution we found for which all angles are simple fractions of $\pi$: $(\pi/3, \pi/3, \pi/3, -\pi/6)$. In this case, $A_1 = A_2 = A_3 = A_4/3$, and $\alpha_1 = \alpha_2 = \alpha_3 = -2\alpha_4$. This class of models, $(\alpha_1, \alpha_1, \alpha_1, \alpha_4)$, with $\alpha_1 = -\cot^{-1} \frac{\cot \alpha_4}{3}$, is of particular interest because the node twists wholly with filament 4, i.e. $\boldsymbol{\omega} = \alpha_4 \hat{\boldsymbol{f}}_4$.

Another simple class of solutions has $(\alpha_1, \pi - \alpha_1, \alpha_3, \pi - \alpha_3)$, in which two pairs of filaments rotate by the same amount in opposite directions. Fig. 10 shows the half-angle of rotation $\omega$ in this case in terms of $\alpha_3$, for a few $\alpha_1$'s. Note that because filaments can be reassigned, $\omega$ is invariant to swapping pairs of filaments. $\alpha_1$ and $\alpha_2 = \pi - \alpha_1$ can be swapped, so $\omega$ is symmetric about $\pi/2$.

The direction of rotation in this restricted model is also interesting. Fig. 11 shows alignments of the node's rotation axis, and directions of the four filaments. The cosine of the angle between the unit vectors, $\hat{\boldsymbol{f}}_i \cdot \hat{\boldsymbol{\omega}}$, lies in the range $\pm\sqrt{2/3} \approx \cos(35.3°)$; this maximum alignment occurs when $\boldsymbol{\omega} \perp$ some other filament. The curves cross when $\alpha_1 = \alpha_3$; in this case, $\boldsymbol{\omega}$ is equally aligned with both vectors, and $\hat{\boldsymbol{f}}_1 \cdot \hat{\boldsymbol{\omega}} = \pm 1/\sqrt{6}$. In this restricted model, $\cos^{-1}(\hat{\boldsymbol{f}}_1 \cdot \hat{\boldsymbol{\omega}}/\sqrt{2/3})$ goes through $(\alpha_3 = 0, \pi/2)$, $(\alpha_3 = \alpha_1, \pi/4)$, and $(\alpha_3 = \pi/2, 0)$, with similar constraints for other $\hat{\boldsymbol{f}}_i$'s. The curves with $\alpha_1 = \pi/4$ are entirely straight. For other $\alpha_1$'s, we checked that the curves closely track hyperbolas satisfying these three constraints.

The node's rotation half-angle $\omega = \pi/2$ whenever any $\alpha_i = 0$ (or $\alpha_i = \pi$). This means that the node rotates by $\pi$, the same as a reflection through some plane. But the parity inversion cancels this reflection along that plane. This is consistent with the lack of rotation along axis $i$ with $\alpha_i = 0$ or $\pi$, leaving the vertices attached to the filament.

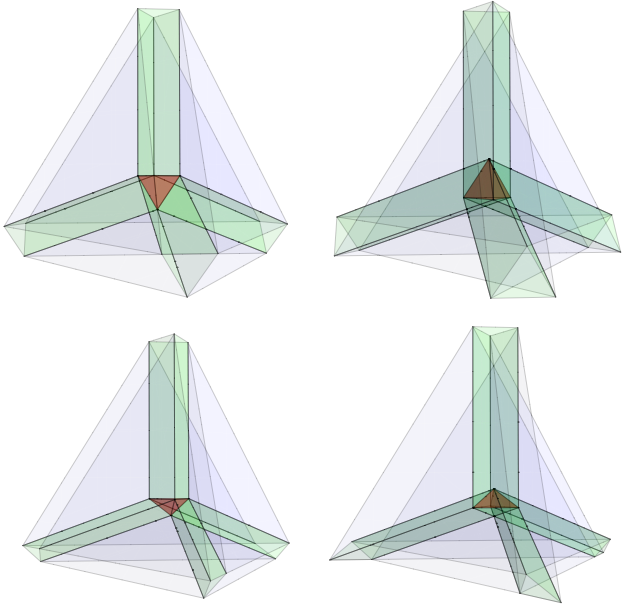

**Figure 7.** Caustic locations in tetrahedral-collapse models. Filament creases, in green, are indicated by triangular tubes, intersecting at the central node. Wall creases, in blue, extend from filament edges through the thin lines drawn between filaments. **Left:** Pre-folding/collapse (Lagrangian). **Right:** Post-folding/collapse (Eulerian). **Top**: An irrotational model ($\alpha_1 = \pi/2$). Each filament vector $\hat{\boldsymbol{f}}_i \perp$ a face of the central tetrahedron. Walls, filaments, and the node invert along their central planes, axes, and point, but remain connected as before. Void regions simply move inward. All 15 initial regions overlap at the center. **Bottom**: A rotational model ($\alpha_1 = \pi/6$). The top filament rotates counter-clockwise by $\pi/3$, while the smaller, bottom filaments rotate clockwise by $2\pi/3$. Views down the top filament appear in Figs. 8 and 9, which can also be clicked to give a 3D figure.

A numerical search verified that the only models with $\omega = 0$ have all $\alpha_i = \pi/2$. As shown in Fig. 13, this seems to be the case in irregular models as well, although there are areas of irregular-model parameter space that were not explored.

That is, it seems that *rotation in the node happens if and only if there is rotation in filaments*. Furthermore, rotation in one filament must accompany rotation in at least one other.

A key missing piece for future investigation is an analytic equation relating the node's rotation vector $\boldsymbol{\omega}$, and its other properties. $\boldsymbol{\omega}$ is probably the most important quantity in the model, given its possible observational relevance (as a galactic rotation). The parameter space of the angles $\alpha_i$ (and filament directions $\hat{\boldsymbol{f}}_i$) and solution space of $\boldsymbol{\omega}$ are both high-dimensional, so it is difficult to gain intuition into how all of the quantities relate.

Interestingly, even within the constraints of the regular-tetrahedral filament arrangement, the node generally becomes more squashed as the spin increases, suggestive of the flattened nature of a spinning disk. However, it is not obvious how to quantify flattening in a general tetrahedron. If a protohalo collapses with a somewhat tetrahedral shape, after virialization, the halo would likely become ellipsoidal,

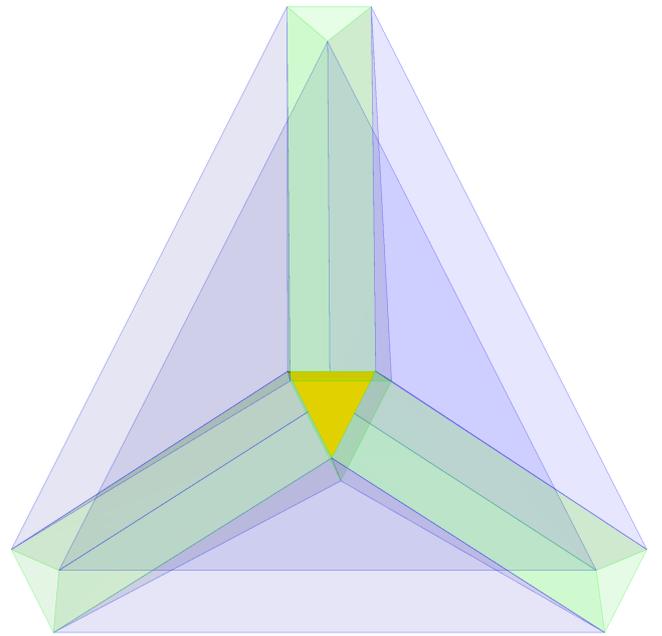

(Lagrangian space)

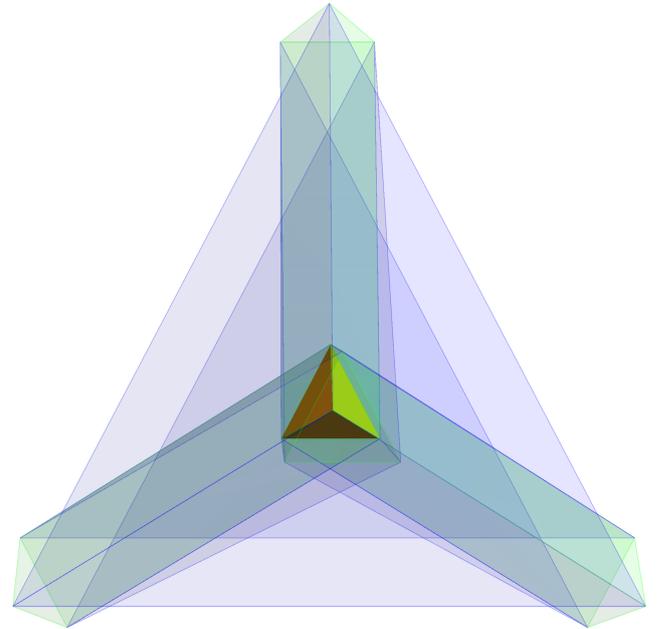

(Eulerian space)

**Figure 8.** Surfaces showing the positions of caustics before and after collapse, in a regular-tetrahedral, irrotational model with all filament rotation angles $\alpha_i = \pi/2$. Surfaces are shown cropped from their infinite extent. The yellow tetrahedron represents node caustics, green triangular tubes represent filament caustics, and parallel pairs of blue planes represent wall caustics. Click in a 3D-capable PDF reader for an interactive figure. Click in a 3D-capable PDF reader for an interactive figure. Right-click for a menu of viewing options, including orthographic projection and transparency. Buttons rotate the model to look down each of the 4 filaments. See Supplemental Material for an animation showing the motion of the caustics from Lagrangian to Eulerian space, also at http://skysrv.pha.jhu.edu/~neyrinck/irrotet.mov.

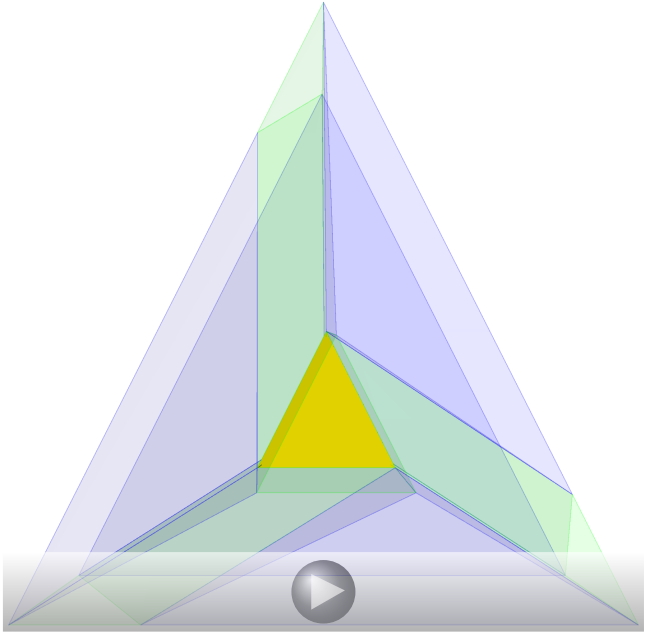

(Lagrangian space)

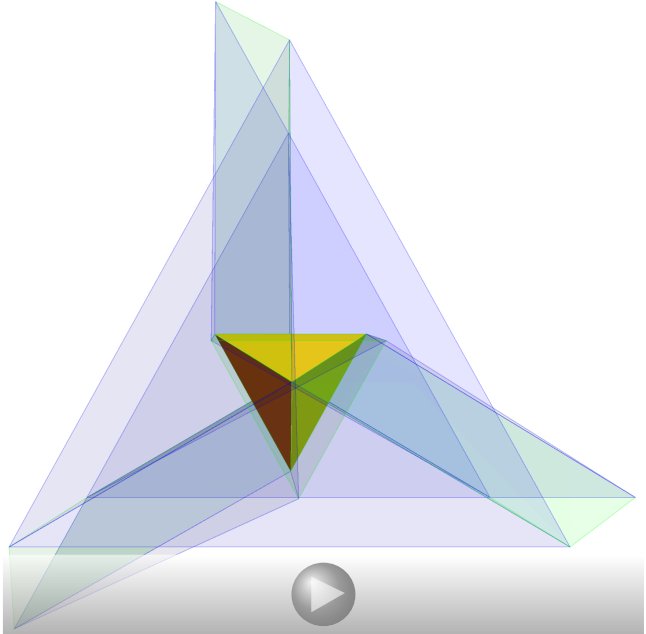

(Eulerian space)

**Figure 9.** Same as Fig. 8, but with filament rotation angles $\alpha_1 = \alpha_2 = \alpha_3 = \pi/3$ and $\alpha_4 = 5\pi/6$, or $1 - 5\pi/6 = \pi/6$ in the opposite direction from the other three. Filaments point in directions of regular-tetrahedron vertices, but the tetrahedral node is irregular. Filament cross-sectional areas $A_1 = A_2 = A_3 = A_4/3$. Click in a 3D-capable PDF reader for an interactive figure. Right-click for a menu of viewing options, including orthographic projection and transparency. The buttons rotate the model so that the camera looks down each of the 4 filaments. See Supplemental Material for an animation showing the motion of the caustics from Lagrangian to Eulerian space, also at http://skysrv.pha.jhu.edu/~neyrinck/rotet.mov.

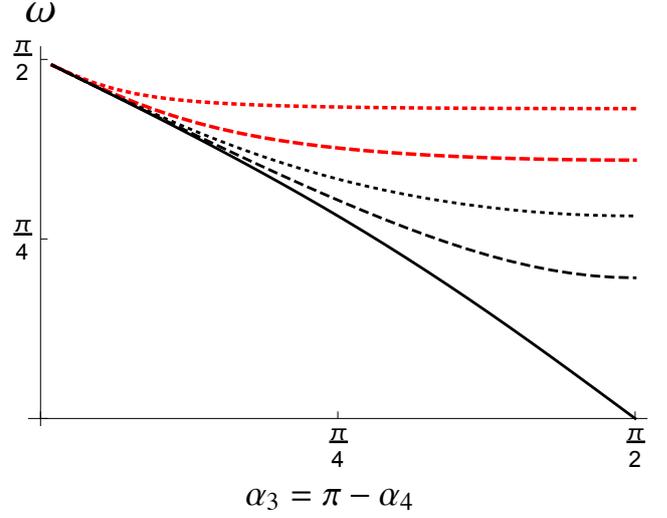

**Figure 10.** The angle of rotation $\omega = |\boldsymbol{\omega}|$ in a model where the node rotation angles $\alpha_2 = \pi - \alpha_1$ and $\alpha_3 = \pi - \alpha_4$. $\alpha_1 = \pi - \alpha_2 = \pi/2$ (solid), $\pi/3$ (dashed), $\pi/2$ (dotted), $\pi/6$ (red dashed), and $\pi/12$ (red dotted).

with a shape that might relate to the shape of the initial node. This is also something to investigate further.

### 4.3 Results: Irregular-tetrahedral collapse

The simple laws found above do not generally hold if the filament configuration departs from that of a regular tetrahedron's vertices. Presumably, generalizations of the above laws hold in the irregular case. But the parameter space of irregular-tetrahedral collapse is quite large, so for now, we concentrate on how its properties diverge form the regular-tetrahedral case.

As in the regular case, four filament directions and three filament rotation angles suffice to determine the collapse's properties; we show here how the fourth angle $\alpha_4$ and the node's rotation angle $\omega$ change with the tetrahedron's regularity, in a restricted model with $\alpha_1 = \alpha_2 = \pi/2$.

Fig. 12 shows how $\alpha_4$ changes with $\alpha_3$ in this model for three filament arrangements: regular-tetrahedral (solid), and two irregular arrangements (dashed and dotted), with details given in the captions. In the regular case, the relationship is simply given by the regular-tetrahedral rules: a straight line with $\alpha_4 = \pi - \alpha_3$. In the irregular cases, the curves are altered, but still, $\alpha_4 = \pi$ when $\alpha_3 = 0$, and $\alpha_4 = \pi/2$ when $\alpha_3 = \pi/2$. As in the regular case, it seems that *a single filament cannot rotate by itself*.

Fig. 13 shows how $\omega$ changes with $\alpha_3$, for the same three models. The regular case also appears in Fig. 10. In the irregular cases, the curves depart similarly as in Fig. 12, but with fixed endpoints at $(\alpha_3 = \pi/2, \omega = 0)$ and $(\alpha_3 = 0, \omega = \pi)$. As in the regular case, it seems that *the node rotates if and only if there is rotation in filaments*.

To end this section, we show caustic locations in a rather arbitrary example of a rotational, irregular model, shown in Fig. 14. Here, the filament direction angles $(\theta_2, \theta_3, \theta_4) = (0.5\pi, 0.6\pi, 0.7\pi)$, the filament rotation angles are $\{\alpha_i\} = (\pi/3, \pi/4, \pi/6, -0.110\pi)$, and $\phi_3$ and $\phi_4$ are as in the regular-tetrahedral case. $\alpha_4$, solved-for numerically, departs from its value from Eq. (7) in the regular-tetrahedral case, in which $\alpha_4 \approx -0.093\pi$. The node rotates by an angle $\omega = 0.80\pi$.

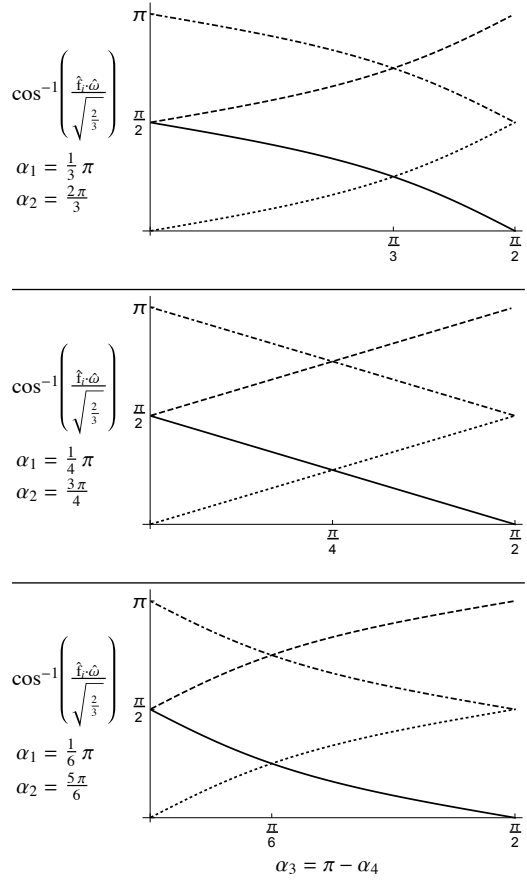

**Figure 11.** Measures of alignments between the unit vector along the node's axis of rotation $\hat{\boldsymbol{\omega}}$ and the four filaments, arranged along vertices of a regular tetrahedron. Solid, dashed, dotted, and dashdotted curves show $\cos^{-1}(\hat{\boldsymbol{f}}_i \cdot \hat{\boldsymbol{\omega}}/\sqrt{2/3})$ for $i = 1$ to 4, respectively. In each plot, $\alpha_1$ is held constant at the value shown, and $\alpha_2 = \pi - \alpha_1$. $\alpha_3$ varies along the x-axis of the plots, and $\alpha_4 = \pi - \alpha_3$. The curves closely follow the shapes of hyperbolas, constrained to go through the endpoints and the intersection points.

## 5 CONCLUSION

This paper describes the tetrahedral collapse model, a toy model of halo collapse that includes the formation of filaments and walls that happen concurrently. Spherical and tetrahedral collapse can be considered as two idealized models of halo formation, simple enough to be tractable without a full simulation, but not expected to happen exactly in nature. Spherical collapse is entirely symmetric, with an idealized initial density profile. Tetrahedral collapse, on the other hand, includes a full description of the geometry of the various cosmic-web elements that form in a more general, anisotropic collapse.

Tetrahedral collapse occurs under the strong 'origami approximation,' which includes a no-stretching condition on the dark-matter sheet (allowing folding in phase space, but not stretching), and a condition that voids are irrotational. No-stretching requires additional caustics (filaments and walls) to form along with nodes (haloes, or protohaloes). Void irrotationality constrains the model further. All four filament densities (per unit length) and spins, and the node's shape and spin, are all determined by the orientation of the filaments, and four other quantities (such as three filament spins, and a scale parameter). In the regular case (with filaments arranged regular-tetrahedrally), we find some intriguingly simple laws, reminiscent of angular-momentum conservation, that relate filament rotations and densities per unit length.

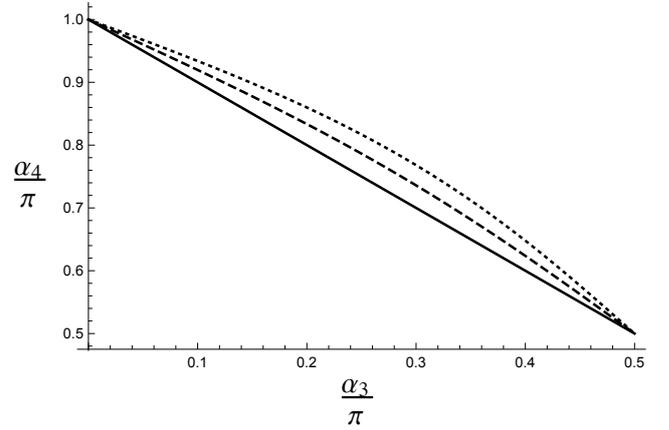

**Figure 12.** In regular and irregular models, the angle of rotation of the remaining filament, $\alpha_4$, in the case that $\alpha_1 = \alpha_2 = \pi/2$. In irregular models, $\alpha_4$ is the same as in the regular case at the endpoints, but it diverges from the regular case between them. The solid curve is the result in the regular case, $\theta_2 = \theta_3 = \theta_4 = \cos^{-1}\frac{-1}{3} \equiv \Delta \approx 0.61\pi$, $\phi_3 = 4\pi/3$, and $\phi_4 = 2\pi/3$. For the dashed curve, $\theta_2$ and $\theta_4$ change: $\theta_2 \to \Delta - \frac{\pi}{10} \approx 0.51\pi$, and $\theta_4 \to \Delta + \frac{\pi}{10} \approx 0.71\pi$. For the dotted curve, there is a further change: $\phi_3 \to 5\pi/4$.

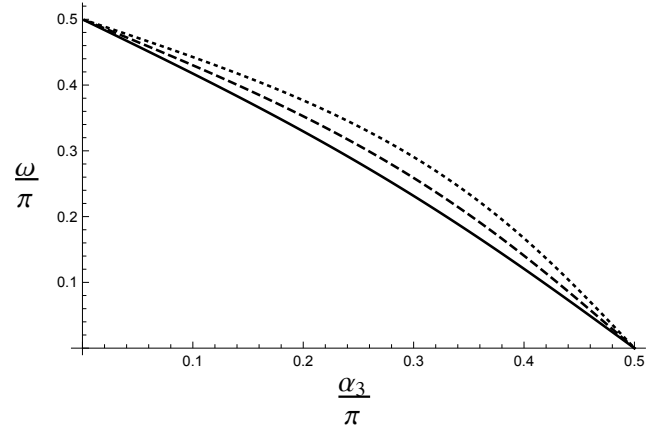

**Figure 13.** Same as Fig. 12, except showing the node's rotation half-angle $\omega$.

Two particular qualitative predictions of the model are that *a filament cannot spin by itself*, and *a halo spins if and only if there is spin in its filaments*. Since haloes gener-ally spin, this implies that *filaments generally spin* as well; filament spin is something that seen little study before. A spinning filament that connects two haloes would tend to correlate their spins, but not perfectly, since the halo spins also depend on its other filaments' spins.

With some assumptions about the magnitudes and distributions of filament spins, which could be measured from simulations, correlations between neighbouring halo spins could be inferred. Indeed, some correlation could possibly be inferred for second- and $n$th-degree neighbours, i.e. haloes joined together, with other haloes in between, although this may get quite complicated. This spin correlation based on connectivity on the cosmic-web graph would be an entirely new approach to this study, complementary to investigation based on the tidal torque theory.

However, it is important not to go too far without verifying predictions of the tetrahedral-collapse model in simulations. A difficulty is that it is not obvious how to measure filament spins. It is already subtle to measure halo spins, which depend somewhat on the choice of halo center and edge. For a filament, choices about a filament's extent and center need to be made all along it, rather than at a single point.

Galaxy intrinsic alignments would likely be the most useful application of the model. Even if its predictions do not hold perfectly, it is likely that aspects could be successfully calibrated with simulations or observations, and it could provide a useful conceptual framework.

Please see `http://skysrv.pha.jhu.edu/~neyrinck/TetCollapseFil2Node/` for an interactive model, using the Wolfram CDF format.

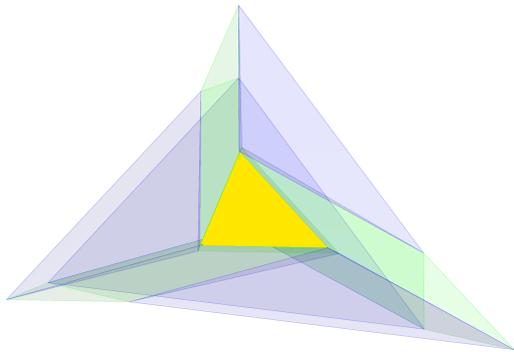

(Lagrangian space)

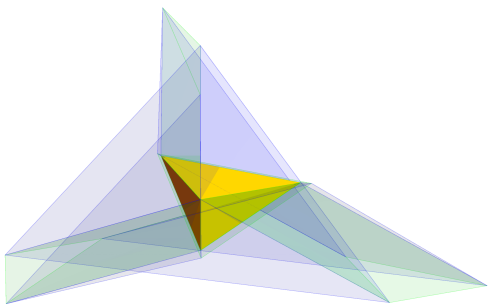

(Eulerian space)

**Figure 14.** Same as Fig. 8, but with filament rotation angles $\alpha_1 = \pi/3$, $\alpha_2 = \pi/4$, $\alpha_3 = \pi/6$, and $\alpha_4 = -\cot^{-1}(\cot\alpha_1 + \cot\alpha_2 + \cot\alpha_3) \approx 0.89\pi$, or $(1 - 0.89)\pi = 0.11\pi$ in the opposite direction from the other three. In this case, the filaments are not in a regular-tetrahedral arrangement, causing filaments to have non-equilateral cross-sections. The ratios of $A_2$, $A_3$ and $A_4$ to $A_1$ are 1.4, 3, and 7.7 to 1. Click in a 3D-capable PDF reader for an interactive figure. Right-click for a menu of viewing options, including orthographic projection and transparency. The buttons rotate the model so that the camera looks down each of the 4 filaments. See Supplemental Material for an animation showing the motion of the caustics from Lagrangian to Eulerian space, also at `http://skysrv.pha.jhu.edu/~neyrinck/irregular.mov`.

**ACKNOWLEDGMENTS**

I thank Robert Lang, Simon White and Stéphane Colombi for helpful conservations, and Robert Lang for permission to use Fig. 4. I am grateful for financial support from a grant in Data-Intensive Science from the Gordon and Betty Moore and Alfred P. Sloan Foundations, and for hospitality at the Institut d'Astrophysique de Paris, where the revision of this paper took place.